\documentclass[pra,aps,superscriptaddress, showpacs, twocolumn]{revtex4}

\usepackage{amsmath}
\usepackage{amssymb}
\usepackage{graphicx}

\newcommand{\ket}[1]{\vert#1\rangle}

\begin{document}

\title{Deterministic generation of arbitrary symmetric states and entanglement classes}

\date{\today}

\author{L. Lamata}
\affiliation{Departamento de Qu\'{\i}mica F\'{\i}sica, Universidad del Pa\'{\i}s Vasco UPV/EHU, Apartado 644, E-48080 Bilbao, Spain}

\author{C. E. L\'opez}
\affiliation{Departamento de F\'{\i}sica, Universidad de Santiago de Chile, USACH, Casilla 307 Correo 2 Santiago, Chile}
\affiliation{Center for the Development of Nanoscience and Nanotechnology, 9170124, Estaci\'on Central, Santiago, Chile}

\author{B. P. Lanyon}
\affiliation{Institut f\"ur Quantenoptik und Quanteninformation, \"Osterreichische Akademie der Wissenschaften, Otto-Hittmair-Platz 1, A-6020 Innsbruck, Austria}

\author{T. Bastin}
\affiliation{Institut de Physique Nucl\'eaire, Atomique et de
Spectroscopie, Universit\'e de Li\`ege, 4000 Li\`ege, Belgium}

\author{J. C. Retamal}
\affiliation{Departamento de F\'{\i}sica, Universidad de Santiago de Chile, USACH, Casilla 307 Correo 2 Santiago, Chile}
\affiliation{Center for the Development of Nanoscience and Nanotechnology, 9170124, Estaci\'on Central, Santiago, Chile}

\author{E. Solano}
\affiliation{Departamento de Qu\'{\i}mica F\'{\i}sica, Universidad del Pa\'{\i}s Vasco UPV/EHU, Apartado 644, E-48080 Bilbao, Spain}
\affiliation{IKERBASQUE, Basque Foundation for Science, Alameda Urquijo 36, 48011 Bilbao, Spain}

\begin{abstract}
We propose a method to generate arbitrary symmetric states of $N$ qubits, which can be easily associated with their entanglement classes. It is particularly suited to quantum optics systems like trapped ions or  superconducting circuits. We encode each qubit in two metastable levels of the system and use a bosonic quantum bus for creating the states. The method is deterministic and relies on a sequence of selective unitary gates upon the qubits within the system coherence time. 
\end{abstract}


\maketitle

\section{Introduction}

Entanglement is one of the most significant features in quantum information science~\cite{NielsenChuang}. Even though a full  characterization of entanglement is still lacking~\cite{HorodeckiReview,GuhneTothReview}, several important steps have been achieved in order to classify entangled states according to different criteria. Among these, the most prominent is the classification under Stochastic Local Operations and Classical Communication (SLOCC). This was firstly obtained for the 3-qubit case~\cite{DurEtAl} and subsequently extended to the 4-qubit case~\cite{VerstraeteEtAl,LamataEtAlMethod,LamataEtAl,Siewert11} and larger numbers of qubits~\cite{Li}. Two $N$-qubit pure states $|\psi\rangle$ and $|\phi\rangle$ are equivalent under SLOCC if and only if $|\psi\rangle=A_1\otimes A_2\otimes...\otimes A_N |\phi \rangle$, where $\{A_i\}_{i=1,...,N}$ are invertible matrices, not necessarily unitary. Two states belong to the same SLOCC entanglement class if and only if they are equivalent under SLOCC. The motivation for this entanglement classification is practical, i.e., two states belonging to the same SLOCC class are able to perform the same quantum information tasks up to a finite probability. Extensions to $N>4$ qubits are cumbersome, due to the fact that the number of entanglement classes becomes infinite already for $N=4$~\cite{DurEtAl}. On the other hand, it is sensible to group the infinite number of classes in a certain set of {\it families}~\cite{VerstraeteEtAl,BastinEtAl}, containing classes that share similar mathematical properties. It is also interesting to analyze entanglement classes and families of states that have specific symmetries. In this respect, the classification for $N$-qubit pure states that are symmetric under qubit permutation has been achieved~\cite{BastinEtAl,MathonetEtAl,Markham12}. We point out that this set of symmetric states contains an infinite number of SLOCC classes in some of the entanglement families. The experimental access to the vast majority of states in the symmetric classes remains an open problem, whose solution would enhance the use of multipartite entanglement in quantum information protocols, e.g., recent studies have shown the particular computational power of nonstandard W-like and GHZ-like symmetric states~\cite{DHondt}.

On the other hand, an increasing interest for operational entanglement classifications, with direct connection to experimental setups, has given rise to proposals based on atoms or ions coupled by spontaneously emitted photons and interference in the detection process~\cite{BastinEtAlExp,BastinWeinfurter}. However, this kind of proposals have a small success probability that decreases exponentially with the number of qubits, restricting their scalable implementation~\cite{DuanMonroe}.
Trapped ions are one of the most controllable quantum systems~\cite{LeibfriedEtAl,Haffner08} and, nevertheless, a limited set of prototypical entangled states such as $W$~\cite{Haffner05} or Greenberger-Horne-Zeilinger (GHZ)~\cite{Leibfried05} states of up to 14 qubits have been achieved~\cite{Monz11}. Proposals for symmetric Dicke state generation, but not their superpositions as required for the much larger set of \emph{arbitrary} symmetric states, have also been put forward~\cite{SolanoMilman,SolanoSantos,SolanoPhonon,CarlosLopez,Retzker07,Vitanov08,SUrabe,Wineland09}. An alternative quantum platform, as superconducting quantum circuits, is the most promising solid-state quantum technology~\cite{WendinShumeiko,WilhelmReview}. Significant steps towards the generation of multipartite entangled states in this implementation have been given~\cite{JohanssonGHZ,Fink09,Bruder10,Martinis10}. However, a protocol for generating the broad set of arbitrary $N$-qubit symmetric states in these or other technologies is still missing. 

Here we propose an efficient protocol for generating arbitrary symmetric $N$-qubit pure states in quantum optics systems like trapped ions or superconducting circuits. Our protocol is deterministic and feasible with current technology, making use of selective interactions to map product states with fixed number of excitations to symmetric Dicke states with the same number of excitations. We also propose a mapping between symmetric entanglement classes and experimental parameters in quantum controllable systems. Finally, we give an alternative and complementary method for generating arbitrary symmetric states based on an initial encoding with a Fock state superposition.

 We first introduce in Section \ref{SectI} the basic idea of our protocol together with a feasibility analysis. In Section \ref{Implementations} we discuss a possible implementation in quantum optics systems. In Section \ref{SectAlter} we give an alternative and complementary method for producing symmetric states based on the generation of an initial Fock state superposition. We give our concluding remarks in Section \ref{SectConclusions}.

 \section{Deterministic generation of symmetric states\label{SectI}}
 
 We consider a set of $N$ qubits that will be coupled with a bosonic mode acting as a quantum bus. The latter can be a motional mode in the trapped-ion case, or a microwave field in the superconducting circuit case. Our aim is to generate an arbitrary symmetric state of the $N$ qubits, which can be expressed as an arbitrary superposition of symmetric Dicke states with variable number of excitations,
\begin{equation}
|\Psi_N\rangle_S=\sum_{k=0}^N c_k |D_{N,k}\rangle,\label{symmetricDicke}
\end{equation}
where  $|D_{N,k}\rangle$ is the symmetric Dicke state with $N$ qubits and $k$ excitations (we will give a 4-qubit example below). We point out that one can directly map the $c_k$ coefficients in this state to the corresponding symmetric entanglement classes through Vieta's formulas~\cite{BastinEtAlExp,BastinEtAl,MathonetEtAl}. We will use this property at the end of the paper to establish a correspondence between entanglement classes for symmetric states and experimental parameters in quantum optics systems.

As the starting point of the  protocol, we prepare the $N$ qubits in states $\ket{g}$. Then, we generate the same superposition as in Eq.~(\ref{symmetricDicke}), but in the basis of states having $k$ last qubits in state $\ket{e}$ and the remaining first $N-k$ qubits in state $\ket{g}$,
\begin{equation}
\mid \Psi_N \rangle= \sum_{k=0}^N c_k \ket{(k)},
\label{psiN}
\end{equation}
where $\ket{(k)}\equiv|g_1 g_2...g_{N-k} e_{N-k+1}...e_N\rangle$.
To illustrate how this task could be implemented let us consider a particular case of a $4$-qubit state. We start from the state with zero excitations $ \ket{\Psi_0}=\ket{gggg}$. Let us apply a local rotation on the fourth qubit such that we have $ \ket{\Psi_1}=\alpha_0\ket{gggg}+\beta_0\ket{ggge}$. Then, we apply a rotation on the third qubit, conditioned to the fourth qubit being in the excited state $\ket{ e}$. The way to do this in terms of elementary gates is shown in Ref.~\cite{NielsenChuang}. This leads to state $ \ket{ \Psi_2}=\alpha_0\ket{ gggg}+\beta_0 \alpha_1 \ket{ ggge}+\beta_0 \beta_1 \ket{ ggee}$. Implementing a similar operation on the second qubit conditioned to the state of the third qubit, and a rotation on the  first qubit conditioned to the state of the second, we finally obtain the desired  four qubit state,
\begin{eqnarray}
| \Psi_4 \rangle &=& \alpha_0\ket{ gggg}+\beta_0 \alpha_1 \ket{ ggge}+\beta_0 \beta_1 \alpha_2 \ket{ ggee} \nonumber\\
&+&\beta_0 \beta_1 \beta_2\alpha_3 \ket{ geee}+\beta_0 \beta_1 \beta_2\beta_3 \ket{ eeee}.
\label{psi4}
\end{eqnarray}
Coefficients $\alpha_i$ and $\beta_j$ from Eq.~(\ref{psi4}) can always be found out of coefficients $c_k$ in Eq.~(\ref{psiN}). Following this procedure, we can generate an arbitrary superposition of states with $k$ excited qubits as defined above. The needed elementary gates grow polynomially in the number of qubits of the target symmetric state. With this example, we have given a specific protocol for generating $| \Psi_4 \rangle$.

  \begin{figure}[h!]
   \includegraphics[width=1\linewidth]{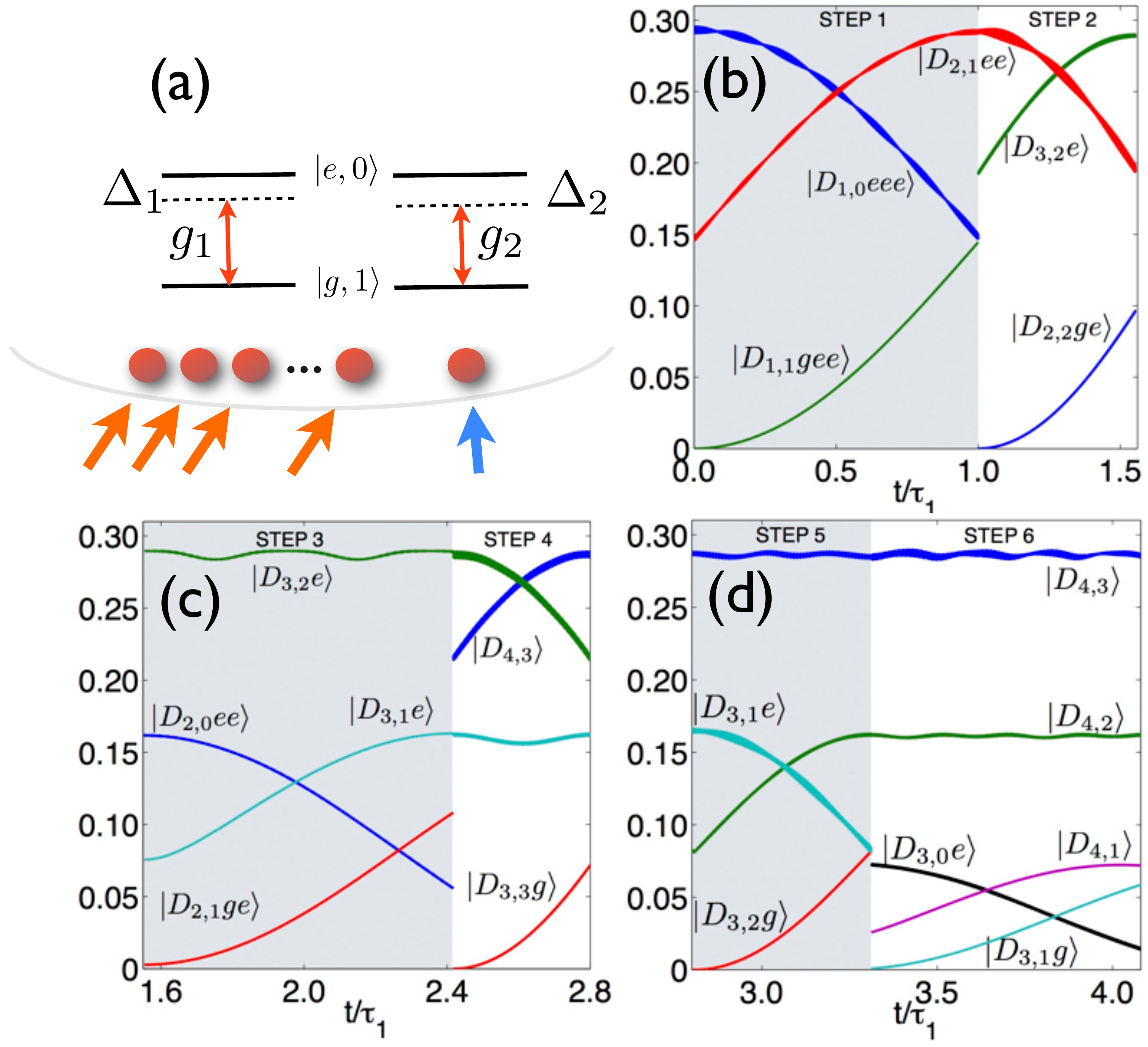}
 \caption{(color online). (a) Scheme of multiqubit system as proposed in the text. In one step of the proposed protocol, the first $n$ qubits are addressed with a global red sideband interaction of Rabi frequency $g_1$ and detuning $\Delta_1$, while the last qubit (first one from the right in the figure) is addressed with a second red sideband driving of Rabi frequency $g_2$ and detuning $\Delta_2$. Here $|e,0\rangle$ (resp., $|g,1\rangle$) is the excited internal state with no excitations in the bosonic mode (resp., ground internal state with one excitation in the bosonic mode). (b-d) Probabilities for the evolved state of being in the corresponding states shown in the curves, as a function of time and the respective step, in the protocol for obtaining  state $|\Psi_4\rangle_S$ in Eq.~(\ref{symmetricDicke}), with $c_k=\{1,2,3,4,5\}/\sqrt{55}$. Here $\tau_1= \pi / (4 \beta)$. Depicted is the time evolution of the probabilities for the three states appearing in Eq. (\ref{Dnk}), at each step: $|D_{N,k}\rangle$, $|D_{N-1,k-1}\rangle |e\rangle$ and $|D_{N-1,k}\rangle|g\rangle$.}\label{Fig1}
  \end{figure}

The next step in the protocol is to coherently map the states with $k$ excitations to symmetric  Dicke states $|D_{N,k}\rangle$ of $N$ qubits with $k$ excitations. As a first strategy  to undertake this task, we observe that a symmetric Dicke state with $k$ excitations can always be decomposed as follows,
\begin{eqnarray}
\ket{D_{N,k}} & = & \frac{d_{N-1,k-1}}{d_{N,k}}\ket{D_{N-1,k-1}}\ket{e}\nonumber\\ &&+\frac{d_{N-1,k}}{d_{N,k}}\ket{D_{N-1,k}}\ket{g},
\label{Dnk}
\end{eqnarray}
where $d_{N,k}$ is the normalization of a symmetric Dicke state of $N$ particles with $k$ excitations, e.g., $|D_{N,1}\rangle=(|eg...g\rangle+|geg...g\rangle+...+|g...ge\rangle)/d_{N,1}$. From this expression, we learn that a symmetric Dicke state of $N$ particles with $k$ excitations can be obtained by symmetrically adding one excitation to a symmetric Dicke state of $N-1$ particles with $k-1$ excitations. Thus, if we are allowed to realize a Rabi flopping between $\ket{D_{N-1,k-1}}\ket{e}$ and $\ket{D_{N-1,k}}\ket{g}$, we can obtain a Dicke state of $N$ particles with $k$ excitations. Such Rabi flopping could be realized by an interaction that annihilates one excitation in one qubit and distributes it symmetrically in a set of qubits. This can be achieved by coupling a set of $n$ qubits with a bosonic quantum bus by means of a red-sideband interaction with a detuning $\Delta_1$ and a collective Rabi frequency $g_1$. Additionally, we need another qubit that is coupled with a second red-sideband driving with a detuning $\Delta_2$ and Rabi frequency $g_2$, see Fig.~\ref{Fig1}a. Provided that $\Delta_1, \Delta_2 \gg g_1, g_2$, the system is described by the Hamiltonian
\begin{equation}
H=\hbar\lambda_1 \sum_{i\ne j}^{n}\sigma_i \sigma_j^{\dagger}+\hbar\sum_{i=1}^{n}(\beta \sigma_i\sigma^{\dagger}e^{-i\delta t}+\beta^{*} \sigma_i^{\dagger}\sigma e^{i\delta t}),
\label{H}
\end{equation}
where $\sigma_i$ is the spin ladder operator associated to qubit $i$, and $\sigma$ is the respective ladder operator associated to the additional qubit, $\delta=\Delta_1-\Delta_2-\lambda_2+\lambda_1$, $\beta=g_1^{*}g_{2}/{\bar{\Delta}}$ and $\lambda_j=|g_j|^{2}/\Delta_j$, with $1/\bar{\Delta}=(1/2)(1/\Delta_1+1/\Delta_2)$. The main advantage of this Hamiltonian is that it allows independent evolution inside the subspaces $\{\ket{D_{n,k-1}}\ket{e}, \ket{D_{n,k}}\ket{g}\}$, and it permits to selectively choose which subspace will evolve, provided by an appropriate  tuning of the parameter $\delta$. The selectivity of a given subspace is possible when $\lambda_1\gg\beta$. The mapping of the $N$-qubit basis, as defined in state (\ref{psiN}), can be understood from our example for the four qubit state  (\ref{psi4}).  As shown in the following diagram,
\begin{equation}
  \begin{array}{ccccc}
    \ket{D_{4,0}}&\ket{D_{3,0}e} & \ket{D_{2,0}ee} & \ket{D_{1,0}eee} & \ket{D_{4,4}} \\ 
    ¥ & ¥ & ¥ &_{1} \downarrow & ¥   \\ 
    ¥ & ¥ & ¥ & \ket{D_{2,1}ee} & ¥\\ 
    ¥ & ¥ &_3 \downarrow&_2 \downarrow & ¥    \\ 
    ¥ &  &  \ket{D_{3,1}e} &  \ket{D_{3,2}e}& ¥ \\ 
    ¥ & _6\downarrow& _5\downarrow& _4\downarrow & ¥     \\ 
    \ket{D_{4,0}} &\ket{D_{4,1}} & \ket{D_{4,2}} & \ket{D_{4,3}} &\ket{D_{4,4}}, \\ 
  \end{array}
  \label{table1}
  \end{equation}
 we access the first two qubits in such a way that, under the dynamics given by Hamiltonian (\ref{H}), the system only evolves within the subspace $\{\ket{D_{1,0}eee},\ket{D_{1,1}gee}\}$.  Thus, after an evolution time such that the corresponding amplitudes in  (\ref{Dnk}) are satisfied, we obtain state $\ket{D_{2,1}ee}$. In steps (ii) and (iii), $\delta$ is chosen in such a way that only  $\ket{D_{2,1}ee}$ or  $\ket{D_{2,0}ee}$ will respectively evolve  to states $\ket{D_{3,2}e}$ and $\ket{D_{3,1}e}$. In Table \ref{table2}, the different parameters for each step are shown. We also include the fidelities for the respective steps obtained performing numerical simulations of the dynamics, for obtaining state $|\Psi_4\rangle_S$ in Eq.~(\ref{symmetricDicke}), with $c_k=\{1,2,3,4,5\}/\sqrt{55}$. Assuming that the fidelity of obtaining the initial state (\ref{psi4}) is close to 1, the total fidelity for generating state $|\Psi_4\rangle_S$ is $\sim 0.989$. The fidelity loss is due to the dispersive character of the interaction in Eq.~(\ref{H}) and imperfections in the selectivity condition. The probabilities of the dynamics of the process can be seen in Figs.~\ref{Fig1}b, 1c, and 1d. 
 
  \begin{table}
\begin{center}\begin{tabular}{|c|c|c|c|c|c|}\hline step & $\delta$ & $g_2$ & $\Delta_2 $ & $\Delta_1$ & Fidelity \\\hline 1 & $0$ & 0.1 & 20.0495 & 20 & 0.999 \\ \hline 2 & $\lambda_1 $& 0.1 & 19.9995 & 20 & 0.995 \\\hline 3 & $  -\lambda_1 $ & 0.1 & 20.0995 & 20 & 0.992 \\\hline 4 & $2\lambda_1 $& 0.1 & 19.9495 & 20 & 0.989 \\\hline 5 & $ 0 $& 0.1 & 20.0495 & 20 & 0.987 \\\hline 6& $-2 \lambda_1 $& 0.1 & 20.1495 & 20 & 0.989 \\\hline
\end{tabular}  \caption{\label{table2} Parameters for the preparation of state $|\Psi_4\rangle_S$ in Eq. (\ref{symmetricDicke}), with $c_k=\{1,2,3,4,5\}/\sqrt{55}$, in units of $g_1$.}

\end{center}
\label{table}
\end{table}

\section{Implementations\label{Implementations}}

Let us now give details of possible implementations. Any physical system that allows for detuned red-sideband interactions among well characterized qubits is appropriate for our proposal. Some examples are trapped ions~\cite{LeibfriedEtAl,Haffner08}, or superconducting qubits~\cite{WendinShumeiko,WilhelmReview}. We consider now a possible trapped ion implementation. The scheme could be implemented, e.g., using a linear string of ions in a Paul trap. To generate a symmetric $N$-qubit entangled state $N$ ions are loaded into the trap. Standard techniques of Doppler and resolved sideband cooling prepare a linear ion crystal, with the axial center-of-mass (COM) motional mode cooled close to the ground state.  The method described to prepare the initial state in Eq.~(\ref{psi4}), and its generalisation, can be done using the universal trapped-ion gate set in Refs.~\cite{Benhelm08,Lanyon11}. To derive an estimate of the four-ion protocol time, we consider singularly ionised calcium ($^{40}$Ca$^+$) ions, which are used by many groups around the world. The ions provide a long-lived ($\sim$ 1 s) quadrupole transition at optical frequencies (729 nm), between the $|S_{1/2}\rangle$ and $|D_{5/2}\rangle$ Zeeman states. A weak static magnetic field is applied to lift the Zeeman degeneracy, providing a large number of transitions that can be used to encode quantum information. With a typical magnetic field of 4.2 Gauss, the splitting is 7 MHz between the $|D_{5/2}\rangle$ levels and 10 MHz between the $|S_{1/2}\rangle$ levels. For a four ion string, Rabi frequencies can typically be tuned anywhere from 0 to about 1 MHz, and red-sideband interactions using the center-of-mass axial mode are straightforwardly implementable. Using $g_1{=}2\pi\times 20$ kHz we calculate a time of 5 ms to prepare the $N=4$ ion Dicke state in Eq.~(\ref{symmetricDicke}) with $c_k=\{1,2,3,4,5\}/\sqrt{55}$. Current trapped-ion technology achieves hundreds of gates per experiment inside a total coherence time of about 30 ms~\cite{Lanyon11}. Our protocol will allow to generate an arbitrary $N=4$ symmetric state in trapped ions with fidelities of about 0.99 inside this coherence time.

In order to detect the final state $|\Psi_4\rangle_S$ after the protocol, one may apply recently developed techniques for symmetric state tomography~\cite{TothSymmetricTomography}. The advantage of these techniques is that the number of observables one has to measure to perform the full tomography grows only polynomially in the number $N$ of qubits. This is due to the symmetry of the considered states under permutation of the qubits. To perform this tomography it is enough to realize a polynomial number of local rotations and $\sigma_z$ measurements upon the different qubits~\cite{TothSymmetricTomography}. The latter can be done in the trapped-ion example via resonance fluorescence with a cycling transition to a metastable excited state~\cite{LeibfriedEtAl,Haffner08}.

We remark that this is a deterministic protocol, i.e., it happens with probability 1, up to experimental imperfections. This should be compared with existing probabilistic proposals and experiments for generating entanglement via photon emission and detection, with success probability for a two-ion entangled state of $\sim 10^{-8}$~\cite{DuanMonroe}. The success probability for $N=4$ ions is, indeed, much smaller in these probabilistic methods given that it decreases exponentially with $N$.

Another appeal of our proposal is the mapping between symmetric entanglement classes and experimental parameters in quantum optics systems. Given that this protocol enables the generation of arbitrary symmetric states of $N$ qubits, we can in particular generate all symmetric entanglement classes~\cite{BastinEtAl}. One may establish a link between each entanglement class and the experimental configuration of the specific implementation. First of all, one establishes a correspondence between the $c_k$ coefficients of a specific $|\Psi_N\rangle_S$ in Eq.~(\ref{symmetricDicke})  and the experimental configuration to generate it. Indeed, the $c_k$ coefficients are already determined by  the set of gates necessary to generate state $|\Psi_N \rangle$ in Eq.~(\ref{psiN}). In our specific example, the set of parameters $\alpha_i,$ $\beta_i$, $i=0,1,2,3$, associated with local gates and conditional rotations~\cite{NielsenChuang}, determines the $c_k$ numbers, and it is attainable, in the trapped-ion example, with the gate toolbox introduced in~\cite{Benhelm08,Lanyon11}. Then, one maps the $c_k$ coefficients to the corresponding entanglement class through Vieta's formulas~\cite{BastinEtAlExp,BastinEtAl,MathonetEtAl}.

\section{Alternative protocol based on initial Fock-state superposition\label{SectAlter}}

Instead of encoding the initial superposition~(\ref{psiN}) of states with different number of excitations with the set of ket vectors $ \ket{(k)}$, one could encode it on the set of Fock states associated with the bosonic mode, $|k\rangle_{pn}$. Thus, this other complementary approach requires the initialization of an arbitrary bosonic state with up to $N$ excitations. In Ref.~\cite{Vitanov08}, it was shown how to create arbitrary Dicke states in trapped ions, and an experiment was performed in Ref.~\cite{SUrabe}. The idea there was to first generate a $k$-phonon motional Fock state of, e.g., the center of mass mode, $|k\rangle_{pn}$, and subsequently to transfer it via adiabatic passage with a red sideband transition to the symmetric Dicke state with $k$ excitations, $|D_{N,k}\rangle$.  Notice that in this way only Dicke states, and not arbitrary symmetric states, are generated. To achieve the latter, we suggest to first initialize the center of mass motional state in an arbitrary superposition of Fock states with up to $N$ excitations, $\sum_k c_k|k\rangle_{pn}$. In order to attain this, we refer to the proposal by Law and Eberly~\cite{LawEberly96} for creating an arbitrary electromagnetic field in cavity QED using Jaynes-Cummings~\cite{JaynesCummings} and carrier interactions. This proposal is directly applicable to trapped ions by substituting the electromagnetic field with the motional center of mass mode. Finally, we propose to apply an adiabatic passage with the red sideband transition, which will map {\it each} phonon Fock state with $k$ excitations to {\it each} symmetric Dicke state with $k$ excitations. This will produce the desired arbitrary superposition of symmetric Dicke states, $\sum_k c_k|D_{N,k}\rangle$, i.e., an arbitrary symmetric state of $N$ qubits. We point out that this is a significant step forward with respect to symmetric Dicke state generation from an entanglement classification point of view. The same idea can be applied to superconducting qubits by substituting the arbitrary motional state by an arbitrary microwave-field Fock-state superposition, which can be, in principle, similarly achieved. 

We estimate the fidelities of our two methods to be similar, e.g., for generating a symmetric Dicke state with 10 ions and 2 excitations, both methods give a fidelity of about 0.98. While for the first method generation of the auxiliary state and subsequent dispersive interactions are needed, the second one requires initialization of the Fock state superposition and subsequent adiabatic evolution for mapping to the Dicke state superposition. Both methods require individual addressing, which is feasible with current technology, e.g., in trapped ions.

\section{Conclusions\label{SectConclusions}}

Summarizing, we have introduced a deterministic protocol for generating arbitrary symmetric states of $N$ qubits, and their respective entanglement classes, by means of selective interactions. This will allow the creation of this kind of states for a few ions or superconducting qubits with high fidelities. We have also given a complementary approach based on the initialization of an arbitrary superposition of Fock states and subsequent mapping to the equivalent Dicke state superposition. This proposal is based on a combination of the Linington-Vitanov~\cite{Vitanov08} and the Law-Eberly~\cite{LawEberly96} methods. We point out that the existence of an efficient symmetric state generation based on single- and two-qubit gates will be highly dependent on the specific available gates in the corresponding implementation. On the other hand, our two protocols are feasible with current technology, are very suitable to available implementations, and are valid for every symmetric state. These proposals will contribute to enhance the sets of states that can be achieved in controllable quantum systems with present day technology.

\section*{Acknowledgments}
 
The authors acknowledge funding from EC Marie Curie IEF and IIF Grants, Basque Government IT472-10, Spanish MINECO FIS2012-36673-C03-02, UPV/EHU UFI 11/55, SOLID, CCQED, PROMISCE, and SCALEQIT European projects, PBCT-CONICYT PSD54, FONDECYT 1121034 and 1100700,
Financiamiento Basal under project FB0807, and Belgian F.R.S.-FNRS.

\end{document}